\documentclass[prl,twocolumn,showpacs]{revtex4}

\newcommand{\sgn}{{\mathrm{sgn}}}

\usepackage{graphicx}
\usepackage{epsfig}
\usepackage{bm}

\begin{document}

\title{Junctions of three quantum wires and the dissipative Hofstadter model}

\author{Claudio Chamon$^{1}$, Masaki Oshikawa$^{2}$
and Ian Affleck$^{1}$}

\affiliation{
$^1$ Department of Physics, Boston University, Boston, MA 02215 \\
$^2$ Department of Physics, Tokyo Institute of Technology,
Oh-okayama, Meguro-ku, Tokyo 152-8551 JAPAN}

\begin{abstract}
  We study a junction of three quantum wires enclosing a magnetic flux. This
  is the simplest problem of a quantum junction between Tomonaga-Luttinger
  liquids in which Fermi statistics enter in a non-trivial way. We present a
  direct connection between this problem and the dissipative Hofstadter
  problem, or quantum Brownian motion in two dimensions in a periodic
  potential and an external magnetic field, which in turn is connected to
  open string theory in a background electromagnetic field. We find
  non-trivial fixed points corresponding to a chiral conductance tensor
  leading to an asymmetric flow of the current.
\end{abstract}
\maketitle

Electric conduction in quantum wires is of much current interest. 
{}From a practical viewpoint, continuing advance in
electronic technology is now reaching a level which requires understanding
electric conduction in the quantum limit.  From a basic science 
perspective, conduction in
low-dimensional systems exhibits many interesting properties due to strong
correlation effects.  A seminal example is discussed by Kane and Fisher
(KF)~\cite{KF}, who predicted that backscattering due to a single impurity,
however small, makes the conductance vanish at low temperature in
the presence of repulsive interactions.  This is a clear manifestation of the
non-Fermi liquid behavior of interacting electrons in one dimension,
which can generally be regarded as a Tomonaga-Luttinger liquid (TLL). A TLL is
nothing but a field theory of free bosons in $1+1$ dimensions, and the effect
of the impurity can be studied as a boundary interaction in the field theory; 
renormalization-group fixed points can be identified with  conformally
invariant boundary conditions.  Study of boundary conditions in conformal
field theory is an active research area in itself, with applications to
string theory and statistical mechanics.

In order to make any useful circuit with quantum wires, one needs a junction
of three or more wires, which generalizes the KF problem.  This turns out to
be a surprisingly rich problem~
\cite{Nayak-etal,Rao,Egger,Lederer,Safi,Yi,Moore} with
a number of open questions, even for three wires. One of the novel aspects of
the multi-lead problem is that, unlike the junction of two wires (KF
problem), electron fermion statistics play a crucial role~\cite{Nayak-etal}.

\begin{figure} 
\epsfxsize=1.in 
\center 
\epsfbox{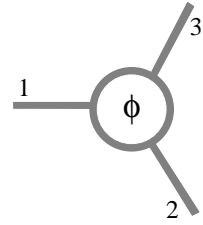} 
\vspace{.2cm}
\caption{A quantum junction of three wires containing a magnetic 
flux $\phi$.}
\label{fig:geom} 
\end{figure}

Here we study a junction of three quantum wires with 
an enclosed magnetic flux $\phi$, as shown in Fig.~\ref{fig:geom}. Using 
a low energy limit Dirac fermion formulation, the physical junction becomes
equivalent to electron hopping between wires, represented by the boundary term
\begin{equation}
H_{\cal B}=-  \sum_{j=1}^3[(\Gamma e^{i \phi/3} \psi^{\dagger}_j \psi_{j-1}
 + \mbox{h.c.})+r\psi^{\dagger}_j\psi_j].
\end{equation}
Here $\psi_j$ is the electron annihilation operator
for the wire $j$ at the junction (with $\psi_0\equiv \psi_3$). 
 $\Gamma$ is the (real) hopping amplitude, $r$ a reflection amplitude 
and $\phi$ is essentially the dimensionless flux through the junction (for  
weak hopping).
$r$ has no effect on the conductance (in the zero frequency, 
zero temperature limit), with the exception of the
non-interacting case,
so we henceforth ignore it. 
{}For simplicity, we consider spinless
fermions, while considering arbitrary strength bulk interactions.  In a TLL,
the interaction strength is essentially contained in a single parameter $g$,
which determines various physical quantities and exponents. $g<1$ and $g>1$
correspond, respectively, to repulsive and attractive interactions, while $g=1$
is the Fermi liquid. 

In the presence of a magnetic flux $\phi$ through the junction, the combination
of the quantum phases due to the flux and fermion statistics is mapped to a
``magnetic field'' for the free boson field at the boundary. Such a theory
has been discussed from the viewpoint of open string theory~\cite{CF}. 
Moreover, the
phase diagram in the presence of both interactions and the ``magnetic field''
at the boundary was studied in a beautiful paper by Callan and Freed
(CF)~\cite{CF}.  We will show that a low-energy renormalization group (RG)
fixed point of our
junction problem is given by the ``magnetic'' boundary condition of CF, which
leads to a chiral conductance tensor corresponding to an asymmetric flow of
the current.
The ``magnetic'' boundary condition exhibits a {\it non-monotonic}
dependence on the TLL interaction parameter $g$, unlike in other applications
of TLL.  In particular, the ``magnetic'' fixed point is stable for $1<g<3$.
This means that an arbitrarily small magnetic flux $\phi$  leads
to a substantial breaking of  time reversal symmetry for this range of $g$.

The electron annihilation operator in wire $j$
is represented as~\cite{KF}
\begin{equation}
  \psi_j \sim \eta_j e^{ i \varphi_j/\sqrt{2}},
\end{equation}
where $\eta_j$ is the so-called Klein factor satisfying
$
 \{ \eta_j, \eta_k \} = 2\delta_{jk}, 
$
which is necessary to ensure the anticommutation relations
of the fermion operators in different wires.
The Klein factors may be represented by the Pauli matrices.
In the low-energy limit, the (imaginary time) 
effective action for the three disconnected 
quantum
wires is given by the three component free boson 
\begin{equation}
S = \int d\tau \; dx \; \sum_{j=1}^3 \frac{g}{4\pi} (\partial_{\mu} \varphi_j)^2 .
\label{eq:S0}\end{equation}
defined over the half-line $x>0$, with the junction at $x=0$.  
Neumann boundary conditions $\partial \varphi_j /
\partial x =0$ are satisfied. 

The boundary interactions lead to renormalization to an infrared 
fixed point with a different boundary condition.  However,  
current conservation at the junction requires the ``center of mass'' field
\begin{equation}
\Phi_0 = \frac{1}{\sqrt{3}} (\varphi_1 + \varphi_2 + \varphi_3)
\end{equation}
to always obey  Neumann boundary conditions.
Thus the remaining degrees of freedom at the boundary
comprise a two component  boson field 
$\vec{\Phi}=(\Phi_1,\Phi_2)$ defined by
$\Phi_1 = (\varphi_1 - \varphi_2)/\sqrt{2}$, 
$\Phi_2 = (\varphi_1 + \varphi_2 - 2 \varphi_3)/\sqrt{6}$.
Using this new basis, the hopping term is written as
\begin{equation}
S_{\cal B}=
i\Gamma e^{i\phi/3}
\int d\tau \sum_{a=1}^3 \eta_a \;e^{i \vec{K}_a \cdot \vec{\Phi}}
+ \mbox{h.c.},
\label{eq:SB}\end{equation}
where $\vec{K}_1 = (-1,\sqrt{3})/2,
\vec{K}_2 =  (-1, -\sqrt{3})/2, \vec{K}_3 = (1,0)$.
The scaling dimension of the hopping term in the disconnected limit
($\Gamma=0$) is calculated by standard methods~\cite{KF} as $1/g$,
which is not affected by the Klein factors nor the magnetic flux.
However, if one attempts a perturbation theory in the hopping
amplitude $\Gamma$, a phase factor appears in each term.
 This makes the problem rather different from
the standard free boson field theory, and hence it is difficult
to predict the nature of the strong hopping limit.
In particular, for $g>1$, when the disconnected wire 
fixed point is unstable, it is difficult to identify the
infrared fixed point.

Each order of the expansion of the partition function
contains an integral over the correlation function
\begin{eqnarray}
\lefteqn{\langle
e^{i \vec{L}_1 \cdot \vec{\Phi}(\tau_1)}
e^{i \vec{L}_2 \cdot \vec{\Phi}(\tau_2)} 
\ldots
e^{i \vec{L}_n \cdot \vec{\Phi}(\tau_n)}
\rangle_0 = } \nonumber \\
& \delta_K(\sum_j \vec{L}_j)&
 \exp{[ \frac{1}{g}\sum_{j>k} \vec{L}_j \cdot \vec{L}_k
      \ln{|\tau_j - \tau_k|}^2 ]}.
\label{eq:loop}
\end{eqnarray}
where $\delta_K(\vec{L})=1$ if $\vec{L}=\vec 0$ and $\delta_K(\vec{L})=0$ otherwise,
$\vec{L}_j$ is one of the six vectors $\pm \vec{K}_{1,2,3}$,
and $\langle \rangle_0$ denotes the expectation value
with the Neumann boundary condition on $\vec{\Phi}$.
Thus, identifying the vertex operator $e^{i \vec{L}_j \cdot \vec{\Phi}}$
with the displacement by $\vec{L}_j$ in the two-dimensional plane,
a non-vanishing contribution~(\ref{eq:loop}) corresponds to
a closed loop on a triangular lattice spanned by $\vec{K}_1$ and
$\vec{K}_2$.
The perturbation series is given by the multi-point 
correlation function~(\ref{eq:loop}) to any order, in principle.
However, in the present problem, we pick up extra phase factors
due to the Klein factors (and the flux $\phi$.)
The phase factor for a given loop is given by the product of
phase factors of elementary triangular loops. Referring to 
the two inequivalent triangles in a triangular lattice as ``up'' and ``down'',
the phase factor for the counterclockwise loop on the up-triangle
is determined to be
\begin{equation}
 (-i \eta_3) e^{i \phi/3} (-i \eta_2) e^{i \phi/3} (-i \eta_1) e^{i \phi/3}
 = e^{i \phi}\label{up}
\end{equation}
while that for the counterclockwise loop on the down-triangle is
\begin{equation}
(i \eta_3) e^{- i \phi/3} (i \eta_2) e^{- i \phi/3}
(i \eta_1) e^{- i \phi/3}  = e^{i (\pi - \phi)}.
\label{down}\end{equation}
Namely, loops on the triangular lattice pick
up a phase as if there is a staggered magnetic flux of $\phi$ and
$\pi - \phi$ in each elementary triangle.

Fortunately, we shall see that the above perturbative
expansion coincides with the ``generalized Coulomb gas''
studied by CF~\cite{CF} in the context of the dissipative
Hofstadter model, DHM,  (quantum motion of a single particle under
a magnetic field and a periodic potential, subject to dissipation.)
The ``free'' action reads:
\begin{equation}
S_0[\vec{X}] = \frac{1}{2} \int \frac{d\omega}{(2\pi)^2}
\left[
\alpha |\omega| \delta_{\mu \nu} + \beta \omega \epsilon_{\mu \nu} \right]
X^*_{\mu}(\omega) X_{\nu}(\omega),
\end{equation}
where $\mu,\nu=1,2$ and $\epsilon_{12}=-\epsilon_{21}=1$,
$\epsilon_{11}=\epsilon_{22}=0$. $\alpha$ and $\beta$ are 
related to the dissipation and the magnetic field, respectively.
This determines the propagator:
\begin{eqnarray}
\lefteqn{
 D_{\mu \nu}(\tau) = \langle X_{\mu}(\tau) X_{\nu}(0) \rangle = }
\nonumber \\
&&  - \frac{\alpha}{\alpha^2 + \beta^2}\ln{\tau^2} \;\delta_{\mu \nu}
+ i \pi \frac{\beta}{\alpha^2+\beta^2}\; \sgn{\,\tau} \;\epsilon_{\mu \nu} .
\end{eqnarray}
We now introduce a potential term, which is somewhat different from
the ``rectangular'' one in Ref.~\cite{CF}, as
\begin{equation}
S_V[\vec{X}]
= - V e^{i \delta/3}\int d\tau \sum_{a=1}^3
e^{i \vec{K}_a \cdot \vec{X}} + \mbox{c.c.},
\label{eq:CFpot}
\end{equation}
where $V$ and $\delta$ are chosen to be real.
Expanding the partition function in powers of $V$, a nonvanishing
contribution is given by an integral over the correlation
function
\begin{eqnarray}
\lefteqn{\langle
e^{i \vec{L}_1 \cdot \vec{X}(\tau_1)}
e^{i \vec{L}_2 \cdot \vec{X}(\tau_2)}
\ldots
e^{i \vec{L}_n \cdot \vec{X}(\tau_n)}
\rangle_0 = } \nonumber \\ 
&\delta_K(\sum_j \vec{L}_j) &  \exp{[  \frac{\alpha}{\alpha^2+\beta^2}
    \sum_{j>k} \vec{L}_j \cdot \vec{L}_k
      \ln{|\tau_j - \tau_k|}^2  } \nonumber \\
&&  \!\!\!\!\!\!\!\!
-i \pi \frac{\beta}{\alpha^2+\beta^2} \sum_{j>k}\vec{L}_j \times \vec{L}_k
\;\sgn(\tau_j - \tau_k) ],
\label{eq:loopCF}
\end{eqnarray}
where $\vec{L}_j$ is again one of the six vectors $\pm \vec{K}_{1,2,3}$.

The perturbation series in the DHM can be 
made to match that in  the quantum wire model, including
the Klein factors and flux, for appropriate choice of $\alpha$ and $\beta$.
The  absolute value of the correlation functions agree exactly if
\begin{equation}
  \frac{\alpha}{\alpha^2+\beta^2} = \frac{1}{g} .
\label{eq:alpha}\end{equation}
The phase factor for any loop in the DHM
is determined by that for two elementary triangles:
\begin{equation}
  \pi \frac{\beta}{\alpha^2 + \beta^2} \vec{K}_1 \times \vec{K}_2 \pm \delta
   = \frac{\sqrt{3} \pi}{2} \frac{\beta}{\alpha^2 + \beta^2} \pm \delta,
\label{eq:beta}\end{equation}
where $+$ and $-$ signs apply to the up and down triangles, just 
as in the quantum wire problem,  (\ref{up}) and (\ref{down}).
Thus, if one chooses
\begin{equation}
\sqrt{3} \pi \frac{\beta}{\alpha^2 + \beta^2} = (2n - 1) \pi
\label{eq:n}\end{equation}
with an integer $n$, there is an appropriate $\delta$ which makes
the two expansions based on~(\ref{eq:loop})-(\ref{down}) and~(\ref{eq:loopCF})
coincide exactly including the phases.
Because the phases are only defined modulo $2\pi$, there is
actually an infinite number of different choices of $\alpha$ and 
$\beta$ labeled
by the integer $n$.
Each choice defines a quite different theory with respect to the
dynamical variable $\vec{X}$, but gives the identical generalized Coulomb gas.

Now let us consider the case of $g>1$.
Then, as we have already discussed, the electron hopping is
a relevant perturbation at the disconnected limit. In the 
DHM, for any choice of $n$, $V$ is a relevant perturbation 
for $g>1$. We would like to find the infrared stable 
fixed point reached in the low energy limit. A simple 
guess is that it occurs  at $V\to \infty$.
The stability of the $V\to \infty$ fixed point 
can be determined using the instanton method~\cite{CF}.
Namely, in the strong potential limit, the $\vec{X}$ field is pinned
at one of the minima of the potential (\ref{eq:CFpot}).
The leading perturbation to this limit is given by a tunneling between
the neighboring minima, represented by an ``instanton.''
A calculation following Ref.~\cite{CF} gives the scaling dimension of
such perturbation to be $\alpha$ times the distance between the nearest
minima squared where $\alpha$ is determined by 
(\ref{eq:alpha})-(\ref{eq:n}). 
For a generic value of $\phi$, minima of the potential~ (\ref{eq:CFpot})
form a triangular lattice with lattice constant $2/\sqrt{3}$.
Thus the dimension of the most relevant operator at this $V\to \infty$ 
fixed point is:
\begin{equation}
\Delta_n=\frac{4g}{3+(2n-1)^2g^2},
\label{eq:dimchi}
\end{equation}
This implies that the $V\to \infty$ fixed point is unstable for 
any value of $g$ for all choices of $n$ except for $n=0$ or $1$ where 
it is stable for $1<g<3$. In this range of $g$ we expect 
that the infrared limit of the junction model corresponds to 
the $V\to\infty$ limit of the DHM for $n=0$ or $1$.
The infrared fixed point for other values of $n$ must correspond to
nontrivial intermediate $V$ fixed points, for which we do not know
explicit solutions.
The infrared fixed point for 
each value of $n$ should give the same physical behavior for the junction 
model. In principle we could study this using any choice of $n$.  
However, clearly the $n=0$ or $1$ choices are preferred because only 
in those cases can we explicitly analyze the fixed points. 
Let us denote the RG fixed points corresponding to the strong potential
limit in these cases as $\chi_-$ and $\chi_+$, respectively for $n=0$ and $1$.
Since $\chi_\pm$ have the same stability,
at this stage
we cannot determine which of these two fixed points is realized
in the infrared limit for a generic value of $\phi$.

On the other hand, in the special cases $\phi=\pm \pi/2$
which maximally break
the time reversal symmetry, the choice of the fixed point is unique.
For $\phi=\pi/2$, the choice $n=1$ ($\beta$ positive)
gives $\delta=0$, and the potential
minima forms a triangular lattice as usual.
However the other choice $n=0$ ($\beta$ negative)
gives $\delta=\pi$, for which the
potential minima form a honeycomb lattice with minimum distance $2/3$,
making the strong potential limit unstable for all $g$.
Similarly, for $\phi = - \pi/2$, $n=0$ ($\beta$ negative) is the
unique choice to give a stable fixed point.
This suggests that these fixed points $\chi_{\pm}$
reflect the breaking of time reversal symmetry due to
the magnetic flux $\phi$.
Indeed, the conductance at these fixed points exhibits a chiral
behavior breaking the time reversal invariance, as we show below. 
On the other hand, changing the flux corresponds to
an irrelevant perturbation at these fixed points. 
Thus we conjecture that the flow from small $V$ goes to the 
$V\to \infty$ fixed point in the $n=1$ representation, 
but not in the $n=0$ representation, where we expect the 
flow to be to a non-trivial fixed point, for $0<\phi <\pi$.
Conversely, for $-\pi <\phi <0$, 
we conjecture that the infrared fixed point is given by the 
$V \to \infty$ one
in the $n=0$ representation, not $n=1$.

The conductance at the junction is expressed as a tensor
\begin{equation}
  I_j = \sum_k G_{j k} V_k ,
\end{equation}
where $I_j$ is the total current flowing into the junction from wire $j$
and $V_k$ is the voltage applied to wire $k$.
{}From current conservation $\sum_j G_{jk}=0$.
It is also clear that the currents depend only on voltage differences.
According to the Kubo formula, the conductance tensor is given by
\begin{equation}
  G_{jk} = -\lim_{\omega \rightarrow 0^+} \frac{1}{\omega}
  \frac{ \partial^2 \ln{Z}}{\partial A_j(-\omega) \partial A_k(\omega)},
\label{eq:Kubo}
\end{equation}
where $\omega$ is the Matsubara frequency, and $A_j$ and $V_j$ are related in
real time $t$ by $d A_j/dt =-V_j$. Assuming that the voltage drops only
across the junction, its effect can be represented as an extra time-dependent
phase factor (vector potential) $ e^{ i (A_j - A_k)}$ in the electron hopping
term from wire $k$ to wire $j$.  It can be conveniently included in the CF
representation as
\begin{equation}
S_V[\vec{X}]
= -V e^{ i \delta/3}\int d\tau \sum_{j=1}^3
e^{i \vec{K}_j \cdot (\vec{X} + \vec{a} ) } + \mbox{c.c.},
\label{eq:CFpotE}
\end{equation}
where $\vec{a} = \sum_{j,k,l}\vec{K}_j \epsilon_{jkl} (A_k - A_l)/3$.
Thus the conductance is related to
\begin{equation}
  \frac{ \partial^2 \ln{Z}}{\partial A_j \partial A_k}
  = \frac{4}{9} \sum_{m,n,\mu,\nu}
  \epsilon_{jm} \epsilon_{kn} K_m^{\mu} K_n^{\nu}
  \frac{ \partial^2 \ln{Z}}{\partial a^{\mu} \partial a^{\nu}},
\label{eq:ZAA}
\end{equation}
where $\epsilon_{12}=\epsilon_{23}=\epsilon_{31}=1$, $\epsilon_{21}
=\epsilon_{32}=\epsilon_{13}=-1$ and $\epsilon_{jj}=0$,
$\mu,\nu=1,2$ refers to the components of two-dimensional vectors
$\vec{a}$ and $\vec{K}_j$.
In fact,
\begin{equation}
  \frac{ \partial^2 \ln{Z}}{\partial a^{\mu}(-\omega ) 
\partial a^{\nu}(\omega )}
  = - \langle \rho_{\mu}(\omega ) \rho_{\nu} (-\omega )\rangle
\label{eq:Zaa}
\end{equation}
is the correlation function of the generalized Coulomb gas density as
defined in Ref.~\cite{CF}, which we use to determine the conductance tensor.
In the strong potential limit of the CF representation,
the $\vec{X}$ field is pinned and the correlation function is given
exactly by
\begin{equation}
    \langle \rho_{\mu}(\omega ) \rho_{\nu} (-\omega )\rangle =
     \alpha |\omega| \delta_{\mu \nu} +\beta \omega \epsilon_{\mu \nu}.
\label{eq:rhorho}
\end{equation}
{}From (\ref{eq:Kubo}) -- (\ref{eq:rhorho}), we obtain the
conductance tensor
\begin{equation}
  G_{jk}^{\pm} = G [(3 \delta_{jk}  -1)
         \pm g \;\epsilon_{jk}]/2,
\label{eq:GDGA}
\end{equation}
with $G = (e^2/h)4g/(3+g^2)$  and $\pm$ corresponding
to the fixed points $\chi_{\pm}$. 
The term proportional to $\epsilon_{jk}$, represents
an asymmetric conduction which is allowed by
the Z$_3$ symmetry of the Hamiltonian once time-reversal
symmetry is broken by the flux.

For example, when a voltage $V_1>0$ is applied to wire $1$ while
$V_2=V_3=0$, the current flowing from wire $1$ to the junction
is $I_1 = G V_1$, which is independent of the asymmetry.
On the other hand, the currents flowing out from the junction to
wires $2$ and $3$ are given by
$-I_2 = G (1 \pm  g) V_1 /2 $ and $-I_3 = G (1 \mp g) V_1/2 $.
In fact, for $g>1$, the current for fixed point $\chi_+$ ($\chi_-$) rather
{\em flows in towards the junction} from wire $3$ (wire $2$),
opposite to the voltage drop,
when the voltage $V_1>0$ is applied
only on wire $1$!

The asymmetric conduction is forbidden under time reversal
but is possible in the presence of
magnetic flux $\phi$.
In the limit of $g\rightarrow 1$, which corresponds to
non-interacting electrons, we find for example
$G_{11}= - G_{12} =1$, $G_{13}=0$ for $\chi_+$.
This asymmetric conduction of non-interacting electrons can be
simply understood as a boundary
condition $ \psi^{out}_{j} = \psi^{in}_{j-1} $.
At $g=1$, there is in fact a continuous family of boundary conditions
characterized by a $3\times 3$ S-matrix $S$ as
$\psi^{out}_{j} = \sum_k S_{jk} \psi^{in}_{k}$ .
Our $\chi_{\pm}$ correspond to special cases of these free fermion boundary
conditions in the $g \rightarrow 1$ limit. 

A remarkable aspect of the fixed points $\chi_\pm$ is
that the conductance~(\ref{eq:GDGA}),
as well as the scaling dimension~(\ref{eq:dimchi}),
exhibits {\em non-monotonic}
dependence on the interaction parameter $g$, unlike in other
known applications of TLL. Making the electron
interaction more attractive can decrease the conductance!

To summarize, we studied a junction of three quantum wires of
interacting spinless electrons with a magnetic flux.
The fermionic statistics, together with the magnetic flux,
bring nontrivial phases into the problem, making it different
from the standard boundary problem of free boson field theory.
It is shown to be equivalent to a certain generalized Coulomb gas
introduced by CF. Using the mapping, we have shown that the chiral fixed points
$\chi_{\pm}$ exhibit an asymmetric conduction and non-monotonic dependence
on the interaction parameter $g$.
The fixed points $\chi_{\pm}$ are stable for $1<g<3$, and
we expect the junction with $\phi=\pi/2$ ($\phi=-\pi/2$)
to be renormalized to the fixed point $\chi_+$ ($\chi_-$),
since it is the unique stable fixed point obtained with
the CF representation at $V\to \infty$.
We conjecture that the system renormalizes to these 
fixed points (for $1<g<3$) for all values of $\phi$
except those which respect time-reversal: $\phi/\pi$ integer.

We note in passing that our results also apply to an equivalent problem. 
If we interpret $\vec \Phi$  
as the co-ordinate of a neutral spin-1/2 
particle,  (\ref{eq:S0}) and (\ref{eq:SB}) 
describe  its  propagation in a periodic 
textured Zeeman field with components 
$B_a(\vec \Phi )=4\Gamma \sin (\vec K_a\cdot \vec \Phi +\phi /3)$, 
 and Hamiltonian  $H_Z=-\vec B (\vec \Phi )\cdot 
\vec S$,  subject 
to dissipation. 

We would like to thank P.~Degiovanni, S. Rao,  
H. Saleur, and X.-G.~Wen for useful
discussions.  This work is supported in part by NSF grants DMR-0203159 (IA)
and DMR-98-76208 (CC), the A.  P. Sloan Foundation (CC), and a Grant-in-Aid
for scientific research from MEXT of Japan (MO).

\end{document}